# Modelling consensus building in Delphi practices for participated transport planning

Michela Le Pira[a]*, Giuseppe Inturri[a], Matteo Ignaccolo[a], Alessandro Pluchino[b]

[a]*Department of Civil Engineering and Architecture (DICAR), University of Catania, Via Santa Sofia 64, 95123 Catania, Italy*
[b]*Department of Physics and Astronomy (DFA), University of Catania, Via Santa Sofia 64, 95123 Catania, Italy*

**Abstract**

In this study a consensus building process based on a combination of Analytic Hierarchy Process (AHP) and Delphi method is presented and applied to the decision-making process about alternative policy measures to promote cycling mobility.
An agent-based model is here used to reproduce the same process of convergence of opinions, with the aim to understand the role of network topology, stakeholder influence and other sensitive variables on the emergence of consensus. It can be a useful tool for decision-makers to guide them in planning effective participation processes.

## 1. The new participatory approach in transport planning

Transport planning is a complex task, mainly because transport problems are often referred to as "wicked" problems with multiple actors and often conflicting interests. Involving stakeholders and citizens from the beginning and all along the planning process is a necessary condition for reaching consensus, while guarantying transparency and pursuing sustainability. The participation process should be planned well in advance, involving the actual stakeholders with appropriate methods (Banister, 2008; Litman, 2009; Cascetta and Pagliara, 2013).

The new approach of participation in decision-making process involves several actors with different roles and it is inspired to some basic concepts, i.e: levels of growing involvement, as represented by the "ladder of citizen participation" (Arnstein, 1969); the main actors involved, i.e. experts, stakeholders and citizens that contribute with different degrees of competence and interest to the decision-making process (Le Pira et al., 2013); the overall transport planning decision-making model (Cascetta et al., 2015), where a "cognitive decision-making" is bounded with stakeholder engagement and quantitative analysis. In the framework of participatory decision-making process in transport planning, planners and experts define the plan structure for the final technical evaluations, stakeholders and citizens are involved in all the planning phases for the definition of objectives, evaluations criteria and alternatives and decision-makers are in charge of the final decision, supported by a performance-based ranking and a consensus-based ranking of plan alternatives (Le Pira et al., 2015b).

In this paper we focus on the group decision-making process that should lead to a consensus-based ranking. To this purpose, Analytic Hierarchy process (AHP) was used to structure the problem and help the evaluation process and a Delphi-type process to promote consensus building. Besides, an agent-based model was used to reproduce the same process and to investigate the phenomenon of consensus building under different scenarios.

The remainder of the paper is organized as follows: section 2 introduces the materials and methods used in the study; section 3 illustrates a case study and its related results; in section 4 the results are discussed and some general conclusions are provided.

## 2. Materials and methods

The study deals with a participation process that involves several phases, with different stakeholders involved and processes adopted. In particular, this paper focuses on the phase of involvement of experts and stakeholders in a combined AHP-Delphi procedure.

Data collected from this experiment are used to derive group preference rankings by different aggregation procedures and to evaluate to what extent interaction contributes to achieve a more shared decision.

Some basic assumptions of the study are the following: (i) it is assumed that an individual expresses his preferences by an ordered list (ranking) of a set of prefixed alternatives; (ii) the ranking of the alternatives can be turned into a binary vector whose components assume the value +1 if the generic alternative A precedes B in the list or –1 if the opposite occurs; (iii) the individual preference rankings must be consistent, i.e. they should derive from logical – non random – judgments; (iv) the collective preference rankings must be transitive, meaning that if alternative A is preferred to B and B to C, then A is preferred to C; (v) the collective preference ranking must be accepted, meaning that it reflects the individual preferences at a reasonable level (or a good degree of consensus).

Based on these premises, in this section the two methods used are presented - AHP to structure the problem and elicit preferences (2.1) and Delphi method to build consensus (2.2) - together with a measure to evaluate the degree of consensus of the collective ranking (2.3). Finally, an agent-based model is presented to simulate the same participation process (2.4). The final aim is to investigate the impact of different scenarios of interaction among stakeholders on the final degree of consensus.

### 2.1. AHP to structure the problem

The Analytic Hierarchy Process (AHP) by Saaty (1980) is based on the representation of a decision-making problem into a tree structured decisions' hierarchy, about the general goal of the plan, sets of specific objectives, evaluation criteria (and possible sub-criteria) and finally plan alternatives aimed at achieving the general goal. A set of pairwise comparison matrices is built by comparing couples of elements at the same level, with respect to the elements of the upper level. The pairwise comparison is made expressing a judgment on a qualitative scale turned into a quantitative one (Saaty, 1987). At each level the pairwise matrices can be transformed into priority vectors with different methods (Saaty and Hu, 1998) and finally a ranking of alternatives can be obtained by combining all the levels.

AHP is widely used in transport planning and management, e.g. to measure the perception of public transport quality (Sivilevičius and Maskeliūnaite, 2010; Mahmoud and Hine, 2013), or for the evaluation of alternatives in transportation planning from a multi-stakeholders multi-objectives perspective (Piantanakulchai and Saengkhao, 2003; De Luca, 2014).

AHP is generally used to elicit single decision-maker opinions, but it can be extended to group decision-making. In the former case, the only condition to respect is judgments' consistency. In the latter case, it is also necessary to define an appropriate procedure to aggregate the individual judgments. There are different prioritization procedures and different aggregation procedures according to which level aggregation is made (Dong et al., 2010): Aggregation of Individual Judgments (AIJ), i.e. the elements of each stakeholder matrix are aggregated into a group matrix, and Aggregation of Individual Priorities (AIP), i.e. a group priority vector is calculated from the individual vectors.

In this paper, the row geometric mean method (RGMM) is used as prioritization procedure and both AIJ and AIP are used to aggregate the individual rankings.

*2.2. Delphi method to build consensus*

The Delphi method (Dalkey and Helmer, 1963) is a procedure that is generally used to make experts' opinions converge on shared solutions. It is addressed to a panel of experts and it is based on some solid assumptions (Pacinelli, 2008), i.e.:

- iterative structure, meaning that participants are called to express their opinions in more rounds;
- anonymity, to avoid bias due to leadership and reciprocal influence of the participants;
- asynchronous communication, with the possibility for the members of the panel to interact remotely and in different times.

At each round of anonymous interaction the members of the panel are asked to align their opinions according to a range where the 50% of the opinions stands (between the first and the third quartiles). The iterations are aimed at mitigating strong positions and finding a collective decision which is shared from the panel.

In principle, it has been used to elicit experts' opinions about the future, with the aim to find "real" values, but it can also be used to explore consensus building in a group.

Being a practice for the convergence of opinions, it can be combined with other methods aimed at eliciting individual preferences.

An interesting approach is the one that combines Delphi practices with multicriteria decision-making methods, such as AHP (Tavana et al., 1993; Vidal et al., 2011) or ANP (Analytic Network Process).

García-Melón et al. (2012) combined the Delphi procedure with ANP to involve stakeholders in a participatory and consensus-building process about sustainable tourism strategies and conclude that, according to the stakeholders involved, this procedure enhanced participation and transparency.

In this study, a Delphi procedure is combined with the AHP method, to elicit preferences of experts and stakeholders about sustainable transport strategies and to see if the anonymous interaction could lead to a convergence of opinions. In this respect, there are multiple ways to measure consensus derived from Delphi, some based on qualitative analysis and others on descriptive statistics (von der Gracht, 2012). With AHP, from the judgments in terms of pairwise comparisons, vectors of preferences on multiple elements are derived. In this case, to measure consensus, we propose the overlap measure as a simple indicator of similarity between two vectors of preferences.

*2.3. Overlap to measure consensus*

Measuring consensus and effectiveness of public participation is a big concern of practitioners. For this purpose, several indicators and models have been proposed to assess "stakeholder satisfaction" (Li et al., 2013) and consensus among experts (Herrera-Viedma et al., 2002).

A simple indicator of consensus has been proposed by the authors to measure the similarity of a collective preference ranking with the individual ones (Le Pira et al., 2015b). If $n$ is the number of alternatives and $m = n(n-1)/2$ is the number of the possible couples of alternatives (i.e. the number of components of each binary vector), the overlap between two lists is defined as:

$$O_{ij} = \frac{1}{m} \sum_{k=1}^{m} V_i^k \cdot V_j^k \qquad (1)$$

where $V_i^k$ and $V_j^k$ are the k-th components of the two binary vectors $V_i$ and $V_j$ representing the preference lists of stakeholders $S_i$ and $S_j$. From this definition follows that $O_{ij} \in [-1;1]$; if $S_i$ and $S_j$ have the same opinion, then $V_i = V_j$ and $O_{ij} = 1$; if all the homologous components $V_i^k$ and $V_j^k$ have opposite signs, then $O_{ij} = -1$; if $V_i$ and $V_j$ are uncorrelated, then $O_{ij} = 0$. The overlap of two stakeholders' opinions about the preference order of $n$ alternatives can be interpreted as the scalar product of the two binary vectors with $m$ components, that is the degree of alignment of two opinions in a $m$-dimensional space. If $N$ is the number of stakeholders involved in the decision, the concept of overlap can be extended to represent the average similarity between the collective list $c$ and the $N$ individual ones:

$$\overline{O}_{i,c} = \frac{1}{N} \cdot \frac{1}{m} \cdot \sum_{i=1}^{N} \sum_{k=1}^{m} V_i^k \cdot V_c^k \qquad (2)$$

The concept of average overlap will be used to measure to what extent stakeholder anonymous interaction due to Delphi method may affect the degree of achieved consensus towards the final decision.

*2.4. Agent-based modelling of Delphi anonymous interaction*

An agent-based model (ABM) is here used (Le Pira et al., 2015a) to reproduce the participatory group decision-making process based on Delphi anonymous interaction. The main aim of the model is to understand what role interaction plays in favouring the convergence of opinions towards a final decision and in escaping from decision deadlock, namely, the so called "Condorcet paradox" or intransitivity paradox. It was studied for the first time in 1785 by the Marquis de Condorcet (1785) who demonstrated that, for a number of alternatives $n>2$, the collective social preference ranking can be intransitive even if the individual preference rankings are transitive. More details about the "Condorcet paradox" in collective decisions and possible ways to escape from it can be found in Le Pira et al. (2015a), Raffaelli and Marsili (2005), Columbu et al. (2008).

In the implemented model, stakeholders are the nodes of a social network and their interactions through the available links is investigated by means of an opinion dynamics model; the consensus formation depends upon critical variables such as the network topology and the degree of interaction. Stakeholders (agents) are endowed with own properties (such as opinion and influence) and act according to simple behavioral rules to reproduce the opinion exchange flows.

The simulation model has been implemented and performed within the software environment NetLogo (http://www.ccl.northwestern.edu/netlogo), particularly suitable for ABM. It consists of several routines, from the creation of the network of stakeholders to the simulation of their opinion exchange until a transitive and shared collective decision is obtained. A list of ordered alternatives is initially randomly assigned to all stakeholders, to represent their individual preferences. Then, a collective list is derived with the so called "Pairwise Majority Rule" (PMR), which is widely used because, in the largest domain, it satisfies all the requirements of a social choice rule (Raffaelli and Marsili, 2005). According to PMR, the collective ranking is obtained by computing how many times each alternative in a pair is preferred to the other one. The pairwise preferences of each individual ranking are coded as components of a binary vector assuming the values of +1 and -1 (e.g., for the couple AB, if A is preferred to B then AB = +1, vice versa AB = -1). Finally, the collective ranking is derived by applying a majority rule to the binary vectors (e.g., if there are five voters and three of them prefer B to A, then we have AB = -1 three times and AB = +1 two times, therefore according to the majority is B>A). The collective ranking derived from PMR can be intransitive, therefore the model aims at investigating how interaction among the agents can help to escape from the deadlock while finding a shared solution.

The interaction process is reproduced by considering that each stakeholder is allowed to know the opinions of his neighbours (i.e. the directly connected nodes in the network) and, at each interaction step, he decides to change his preferences ranking according to the overlap between his ranking and the one of the majority of his neighbors, weighted by their influence. The simulation reproduces a repeated interaction until the average overlap with the collective ranking becomes stable. The final collective ranking is assumed as the transitive "most shared" solution, appreciably reflecting the individual preferences.

In this study, the model is adapted by changing the opinion dynamics in order to reproduce a participation process with the Delphi method. In particular, the network considered is a star, where each node is directly linked only with a "hub" that represents the facilitator of the process, in order to guarantee the condition of anonymity (see section 2.2). At each step of the simulation, the facilitator proposes the collective list to the agents that can decide to change their list according to the similarity with it (i.e. the overlap). Simulation outcomes can give some suggestions about how to manage a Delphi-based participation process and to predict the possible results of interaction.

## 3. Case study

The study is based on a participation experiment where a given number of transport experts and stakeholders were involved to identify policy measures to promote cycling mobility in the city of Catania (Italy).

They were asked to: (i) structure the problem and build the AHP hierarchy, (ii) answer the pairwise comparisons of elements for each level of the hierarchy, (iii) reformulate their judgments after (anonymously) knowing the results of the others. The AHP method was applied - before and after interaction - to derive a priority ranking of alternatives for each actor. The aggregation of the priority vectors was done by using the methods derived from AHP (section 2.1) and the Pairwise Majority Rule (section 2.4). The Delphi method was carried out in two steps of iteration. The panel was composed of seven participants in total: five experts from the University of Catania, with different background (experts on safety issues, land use planning, transport planning), one stakeholder belonging to an association of cyclists and one person that can be considered as a "sophisticated stakeholder" (Mau-Crimmins et al., 2005), being employed in the municipal transport company and being a transport expert.

After a short description of the state of art of cycling mobility in Catania with the structuring of the problem made by the panel, the experiment will be described and the results discussed.

### 3.1. Cycling mobility in Catania

In the last years cycling mobility has been receiving more attention from policy-makers as a sustainable and efficient mode of transport in urban areas. Many cities are adapting to welcome facilities and infrastructures for cyclists, but still lots has to be done, in particular in car-addicted cities.

Catania is a medium-sized city (300,000 inhabitants) located in the eastern part of Sicily, Italy. The city is part of a greater Metropolitan Area (750,000 inhabitants), which includes the main municipality and 26 surrounding urban centers, some of which constitute a whole urban fabric with Catania. The main city contains most of the working activities, mixed with residential areas. Even if several attraction polarities (hospitals, main schools, shopping centers) are spread over the whole territory, the transport demand pattern is mostly radial. Motorized modal split is about 85% private transport and 15% public transport, while the amount of travelled kilometers by bicycle is negligible (even if increasing). Traffic congestion, limited public transport use, little diffusion of cycling and walking for systematic trips, inefficiency of the parking management, absence of city logistics measures are the main critical issues for the transport system of Catania.

Based on these premises, the panel of experts and stakeholders met in a brainstorming session to analyse the problem of promoting cycling mobility in Catania, structuring it into a four-levels hierarchy (Fig. 1). They agreed that four different criteria were necessary to evaluate the alternative measures, i.e. the criteria representing the three dimensions of sustainability (environmental, social and economic) and a transport criterion. The last one was chosen from the panel as an independent criterion even though in principle the transport dimension is encompassed in the other three sustainability criteria. The reason given from the panel is that, in the peculiar case of Catania, an improvement in the transport system contributes itself to a better quality of life.

Seven indicators were chosen to represent the four criteria and then the alternative measures to promote cycling mobility were identified from the panel:

- building a comprehensive cycling network;
- setting up extended 30 km/h zones;
- making information and education campaigns, to increase public awareness towards pros of cycling mobility;
- funding citizens to buy electric bicycles;
- establishing a city bike sharing service.

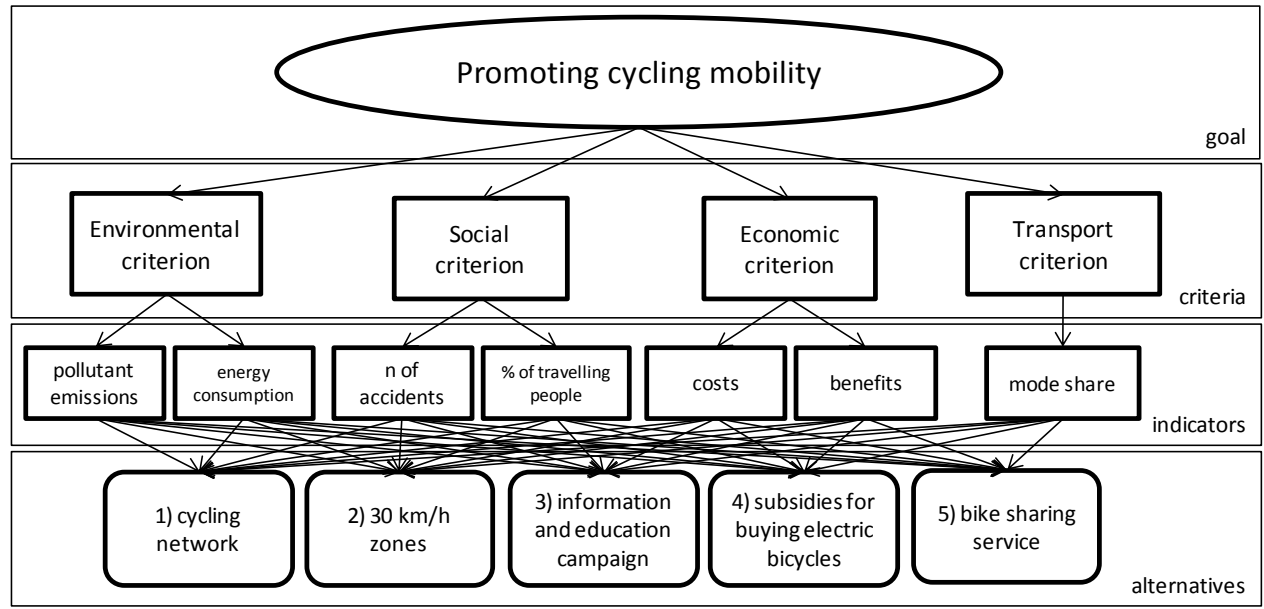

Fig. 1. Hierarchy of the problem "promoting cycling mobility".

*3.2. The Delphi experiment*

The Delphi experiment was conducted in two steps. First, each member of the panel was asked to make judgments in terms of pairwise comparisons between the elements of each level of the hierarchy. Due to the great number of answers they had to give (a total of 79 pairwise comparisons), the facilitator guided them into the whole process paying attention to the consistency of their judgments.

Once all the members filled in the questionnaire, the results derived from AHP were analyzed in order to go on with the second step of the Delphi method. In particular, for each pairwise matrix, the local priority vectors were derived, the first quartile (corresponding to the 25% of judgments) and the third quartile (corresponding to the 75% of judgments) were chosen as reference numbers for the members of the panel to "align" their judgments in the second iteration (see example in Table 1).

By doing this, the results of the second round are in terms of priority vectors for each level of the hierarchy. By aggregating them, the new priority vector for the group is derived, representing the alternative ranking of the panel of experts and stakeholders involved.

Table 1. Example of part of the Delphi questionnaire.

| DELPHI (II round) | Criteria comparison | | | |
|---|---|---|---|---|
| | previous judgment | 1$^{st}$ quartile | 3$^{rd}$ quartile | new judgment |
| environmental criterion | 0.56 | 0.14 | 0.54 | **0.50** |
| social criterion | 0.26 | 0.20 | 0.32 | **0.28** |
| economic criterion | 0.12 | 0.08 | 0.25 | **0.12** |
| transport criterion | 0.06 | 0.12 | 0.26 | **0.10** |

*3.3. Results of the experiment*

The individual preferences were aggregated with the two methods derived from AHP. In particular, in the first round of judgments both the aggregation of individual judgments (AIJ) and the aggregation of individual priorities (AIP) methods were used, while in the second round only the AIP method was used, because the panel was asked to "align" the opinions starting from priority vectors (and not pairwise comparisons). We also aggregated the individual rankings with the Pairwise Majority Rule (PMR) to see if it leads to the same results since it is the aggregation rule used in the agent-based model described in section 2.4.

Table 2 summarizes the results of the average overlap in the first and second round of iteration with the different aggregation methods and sources of individual preferences to be aggregated (from PWC or from priority vectors).

Table 2. Results of collective rankings and average overlap before and after interaction using different aggregation methods.

| Aggregation method | Source of individual preferences | Description | Collective Ranking | | Average Overlap | |
|---|---|---|---|---|---|---|
| | | | I round | II round | I round | II round |
| **AHP-AIJ-gm** | PWC | PWC aggregated into a group matrix (AIJ) using the geometric mean (gm); AHP applied | 2>1>3>5>4 | - | 0.74 | - |
| **AHP-AIP-gm** | Priority vectors | Geometric mean (gm) of the priority vectors derived from AHP for each level (AIP) | 2>1>3>5>4 | 2>1>3>5>4 | 0.74 | 0.80 |
| **PMR** | PWC | PWC transformed in binary vectors (+1 and -1); PMR applied | 2>1>3>5>4 | 2>1>3>5>4 | 0.74 | 0.80 |

The three methods led to the same results (before and after interaction) and the collective ranking resulting from them shows very high values of overlap (0.74 compared with the maximum reachable value of 1). This is due to the fact that the actors involved showed more or less the same level of competence and objectives, even if there are some differences in the individual rankings. After the second round of the Delphi method, there is an increase in the convergence of opinions and this confirms the effectiveness of interaction in the group decision-making process. The opinion changes are expressed in terms of adjustments of the priority vectors derived from AHP (Fig. 2): even though there are not big changes in the weights assigned to the alternatives, in the second round the opinions are closer and less dispersed, with the softening of some initial strong positions.

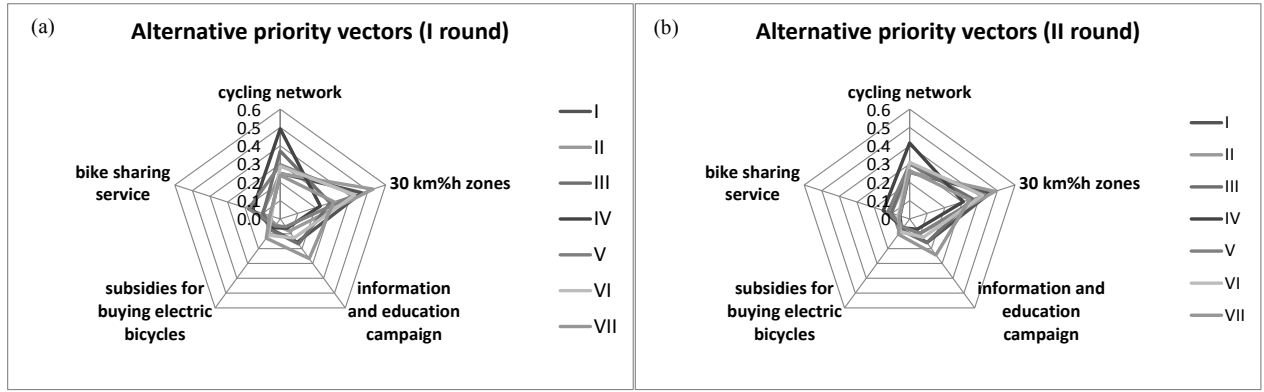

Fig. 2. Alternative priority vectors for the 7 members of the panel in the first round (a) and in the second round (b) of Delphi method.

### 3.4. Results of the agent-based simulations

The same experiment was reproduced with the agent-based model. A network with exactly the same number of agents (7 members of the panel + 1 facilitator) is considered. The network is a "star network", where the hub (i.e. the facilitator) is directly linked with all the others. This can represent quite well the anonymous interaction process among the members of the panel, that could not communicate with each other but knew the other (averaged) answers. Actually, the model is simplified because it starts from their actual preference rankings, evaluates a collective preference ranking through PMR and then agents decide to align their ranking to the collective one according to the similarity with it (in terms of overlap). In other words, while in the real experiment the stakeholders were asked to modify their preferences in terms of weights assigned to the elements of the hierarchy of Fig. 1, in the model they decide to directly change their preference order on alternatives (or maintain it) based on the similarity (i.e. the overlap) with the collective ranking.

Despite its simplicity, results show that the model is able to capture the essence of the phenomenon of consensus building and they are summarized in Table 3.

Several simulations were performed by changing the agents' behaviour in terms of willingness to change their ranking with the collective one. We assume that for each agent the higher the similarity (i.e. the overlap) with the collective ranking the more probable his willingness to change opinion. Therefore, hereafter we will use overlap as a proxy of the willingness to change of the agents.

By modifying the willingness to change of the agents, we found some thresholds that made the outcome changes (see Table 3):
- when it is less than 0.6 a total consensus is reachable (i.e. average overlap = 1);
- when it ranges from 0.6 to 0.8 it is not possible to reach total consensus; in this case the final degree of consensus (i.e. average overlap) is similar to the one obtained with just one step of interaction in the Delphi experiment (0.8 with the Delphi experiment instead of 0.886 with the simulations);
- when it overcomes 0.8 the agents are not willing to change their opinions, therefore they maintain the initial one and the process of consensus building does not start.

The topology of the network was also changed to see how agents' influences can affect the final outcome. All the agents were linked in a fully connected network, the influence was randomly assigned with a Poisson law and the simulations were repeated 100 times to have a statistics of events.

Results show (Table 3) that direct interaction of agents with different degrees of influence is not beneficial in terms of consensus buildings, with lower values of final overlap with respect to the star network, where the anonymous interaction is guaranteed, avoiding the risk of leadership.

Other simulations were performed by assigning initial random preference rankings to the agents: in this case the final degree of consensus is much lower than the one obtained with the data from the experiment, meaning that a homogeneous community of people (in terms of interests and expertise) is much more efficient for the success of a Delphi experience rather than a heterogeneous one. This result suggests that the Delphi method could be more effective in eliciting opinions and finding consensus with an expert group rather than a group of stakeholders with diverging interests.

The model also allows monitoring of possible decision deadlock due to Condorcet paradox (see section 2.4) with random initial conditions: while the star network with few nodes (i.e. 7) makes quite improbable to fall into the paradox because the agents are only linked with the facilitator, an intransitive ranking is more likely to occur in a fully connected network where all the agents influence each other.

Table 3. Results of the agent-based simulations in terms of final average overlap (i.e. degree of consensus).

| Network | | Initial conditions | | | |
|---|---|---|---|---|---|
| | | Delphi experiment | | Random | |
| | | wtc* < 0.60 | wtc = 0.60 ÷ 0.80 | wtc < 0.60 | wtc = 0.60 ÷ 0.80 |
| Star | 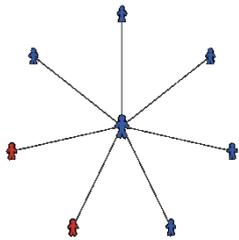 | 1 | 0.886 | 0.425 (Pcycles = 0%) | 0.378 (Pcycles = 0%) |
| Fully connected | 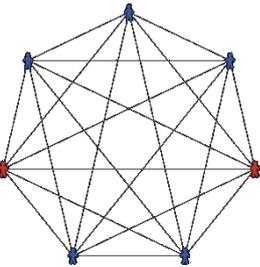 | 0.89 | 0.766 | 0.373 (Pcycles = 15 %) | 0.329 (Pcycles = 7%) |

*wtc = willingness to change

## 4. Discussions and conclusions

The methodology proposed consists of a combined AHP-Delphi method for experts and stakeholders involvement in a group-decision making process oriented to consensus building. An agent-based model was used to reproduce the interaction process where stakeholders are linked in networks and act according to simple behavioral rules on the base of an opinion dynamics model.

Results of the experiment show that the combination of AHP with Delphi method is suitable to support complex group decision-making processes. In this respect, the cooperation of the panel of experts in structuring the problem and sharing criteria, relevant indicators and alternative measures increased the probability to have a good convergence of opinions after only one step of anonymous interaction.

Results of the agent-based model allow to give some suggestions on how to build an effective participation process based on the Delphi method. In particular, it is proved that the outcomes are appreciably influenced by the willingness to change of the agents, resulting in total or partial consensus after interaction. Besides, the guarantee of anonymity avoids the possibility of leadership due to reciprocal influence that can likely reduce the convergence of opinions. Agent-based simulations also suggest that the Delphi method seems to be more effective in terms of consensus building when a quite homogeneous group of experts is involved.

In conclusion, the problem of complex group decision-making processes in transport planning must be tackled with appropriate methods and techniques. In this respect, agent-based simulations of stakeholder interaction can give some useful suggestions to decision-makers and planners in order to support them in designing and guiding an effective participation process.